# This is a pre-print version of the accepted paper!

# Towards optimal quality requirement documentation in agile software development: a multiple case study

Woubshet Behutiye[a], Pilar Rodríguez[b], Markku Oivo[a], Sanja Aaramaa[c], Jari Partanen[d] and Antonin Abhervé[e]

a: M3S Empirical Software Engineering research unit, University of Oulu, Oulu, Finland

b: Universidad Politécnica de Madrid, Madrid, Spain

c: Nokia, Oulu, Finland

d: Bittium Wireless Ltd., Oulu, Finland

e: Softeam, Paris, France

woubshet.behutiye@oulu.fi, pilar.rodriguez@upm.es, markku.oivo@oulu.fi, sanja.aaramaa@nokia.com, jari.partanen@bittium.com and antoin.abherve@softeam.fr

*Abstract*

**Context:** Agile software development (ASD) promotes minimal documentation and often prioritizes functional requirements over quality requirements (QRs). The minimal documentation emphasis may be beneficial in reducing time-to-market for software. However, it can also be a concern, especially with QRs, since they are challenging to specify and document and are crucial for software success. Therefore, understanding how practitioners perceive the importance of QR documentation is valuable because it can provide insight into how they approach this task. It also helps in developing models and guidelines that support the documentation of QRs in ASD, which is a research gap.

**Objective**: We aim to understand practitioners' perceptions of QR documentation and factors influencing this task to derive a model that supports optimal QR documentation in ASD.

**Method**: We conducted a multiple case study involving 12 participants from three cases that apply ASD.

**Results**: Practitioners identify QR documentation as important and perceive it as contributing to ensuring quality, clarifying QRs, and facilitating decision-making. Time constraints, QR awareness, and communication gaps affect QR documentation. Missing and outdated QR documentation may lead to technical debt and a lack of common understanding regarding QRs. We introduce a model to support optimal QR documentation in ASD by focusing on the factors: time constraints, QR awareness, and communication gaps. The model provides a representation and explanation of the factors affecting QR documentation in ASD and identifies mitigation strategies to overcome issues that may occur due to these factors.

**Conclusion**: The study reveals the importance of documenting QRs in ASD. It introduces a model that is based on empirical knowledge of QR documentation practices in ASD. Both practitioners and researchers can potentially benefit from the model. For instance, practitioners can analyze how time constraints or QR awareness affect documentation, see potential issues that may arise from them, and utilize strategies suggested by the model to address these issues. Researchers can learn about QR documentation in ASD and utilize the model to understand the topic. They can also use the study as a baseline to investigate the topic with other cases.

# 1. INTRODUCTION

Agile software development (ASD) has been broadly adopted to meet the demands of dynamic business environments, where requirements change frequently and businesses need to remain competitive (Rodríguez et al., 2012). Consequently, the body of literature on ASD has grown significantly and examines diverse topics, including requirements engineering, adoption challenges and benefits, large-scale adoption, and human and social aspects (Curcio et al., 2018; Dybå and Dingsøyr, 2008; Hoda et al., 2017; Kasauli et al., 2021; Ramesh et al., 2010). Recently, research on the engineering, documentation, and management of quality requirements (QRs) in ASD has drawn considerable interest. QRs, which are also referred as non-functional requirements (NFRs), describe the anticipated quality characteristics of a system to be developed, such as reliability, security, performance, usability and maintainability (Wiegers and Beatty, 2013). They are perceived as difficult to specify and measure (Kitchenham and Pfleeger, 1996) and are determinants of software projects (Glinz, 2007). Some studies have investigated the challenges of managing QRs in ASD (Alsaqaf et al., 2018, 2017; Behutiye et al., 2017; Karhapää et al., 2020), and others have focused on both engineering and managing QRs in ASD (Alsaqaf et al., 2019; Amorndettawin and Senivongse, 2019; Behutiye et al., 2020a; Knauss et al., 2017; López et al., 2017; Oriol et al., 2020). However, QR documentation in ASD remains a research gap that requires more attention (Behutiye et al., 2020a).

In ASD, where documentation is less prioritized (Beck et al., 2001), QRs are often underspecified and undocumented. Consequently, QRs are neglected, resulting in project failures (Ramesh et al., 2010; Sachdeva and Chung, 2017). ASD practitioners are principal stakeholders in the development and management of software. Understanding how they perceive QR documentation is valuable due to QRs' economic implications. For instance, missing QR specifications may incur documentation debt and increase maintenance costs (Behutiye et al., 2020a; Mendes et al., 2016). Moreover, understanding how these practitioners perceive QR documentation can provide insight into their approach to QR documentation. For instance, do practitioners consider documenting QRs to be important, and what are their justifications? How do practitioners perceive various factors, such as time constraints, QR awareness, and communication gaps, which may influence QR documentation? We are interested in these factors since they reportedly affect the documentation and management of QRs in ASD (Behutiye et al., 2020a; Sachdeva and Chung, 2017). Examining such aspects can create knowledge that helps us better understand QR documentation in ASD. We can complement this knowledge in QR documentation with practices from the literature to build models and guidelines that support optimal QR documentation in ASD. This is important since the existing models and guidelines focused on documentation tend to overlook some factors, such as issues that arise from communication gaps, time constraints, and QR awareness, or only focus on specific QRs, such as performance and security. For instance, our documentation guidelines proposal (Behutiye et al., 2017) focuses solely on addressing the limitations of artifacts in ASD for specifying QRs. Others have focused on specifications of performance (Ho et al., 2006) and security (Amorndettawin and Senivongse, 2019; Barbosa and Sampaio, 2015). Our recent review of the literature regarding QR management in ASD (Behutiye et al., 2020a) reveals that most of the existing strategies for managing QRs focus on addressing the limitations of ASD in handling QRs and the neglect of QRs. These strategies address the limitations of user stories in specifying and documenting QRs. Nevertheless, QR documentation and management strategies, specifically tools, models, and guidelines, that fit the short iteration cycles of ASD and address communication gaps regarding QRs and challenges in QR awareness are scarce (Behutiye et al., 2020a). According to Voigt et al. (2016), strategies are needed to support documentation in ASD.

In the ASD literature, some studies have investigated documentation and QRs (Behutiye et al., 2017; Hoda et al., 2012; Kopczyńska et al., 2020; Stettina and Heijstek, 2011; Voigt et al., 2016). However, except for our previous work (Behutiye et al., 2020b), we could not find any studies investigating practitioners' perceptions on the importance of documenting QRs. Based on this gap and the need for models and guidelines to support documentation and management of QRs in ASD (Behutiye et al., 2020a), we conducted this study. This study is conducted in the context of the Q-Rapids project[1], which was an EU horizon 2020 project aimed at defining quality aware rapid software development framework. The paper extends our initial work published in Euromicro SEAA 2020 (Behutiye et al., 2020b). We introduce an initial version of a model that has the ultimate goal of providing support for optimal QR documentation in ASD by considering various factors, such as time constraints, QR awareness, and communication gaps among team members. The model suggests strategies to mitigate challenges encountered in documenting QRs and support practitioners on documenting QRs in ASD. We also examine practitioners' perceptions of the importance of documenting QRs, the factors that may affect QR documentation (i.e., time constraints, QR awareness, and communication gaps among team members), and their perceptions of the consequences of missing and outdated QR documentation. Therefore, our study answers the following research questions:

- RQ1. How do practitioners perceive the importance of documenting QRs in ASD?
- RQ2. How do practitioners perceive factors that may influence QR documentation in ASD?
- RQ3: What are the consequences of missing or outdated QR documentation in ASD?
- RQ4. How can we support optimal QR documentation in ASD?

---

[1] https://Q-Rapids.eu

In answering RQ1, we investigated whether practitioners consider documenting QRs in ASD to be important or not and collect data regarding their justifications. To answer RQ2, we collected practitioners' feedback on how they perceive factors that may influence documenting QRs in ASD. We collected their perceptions on how time constraints, QR awareness, and communication gaps regarding QRs among team members affect QR documentation. In addressing RQ3, we collected evidence regarding the possible consequences of missing and outdated QR documentation in ASD. To answer RQ4, we aim to derive a model to support optimal QR documentation in ASD.

We found that practitioners perceive QR documentation as important for various reasons, such as ensuring quality, clarifying QRs, and assisting decision-making. ASD teams may tailor their documentation practices to fit the sprint duration (e.g., when working in short sprint durations, they allocate sprint dedicated for handling QR documentation). Practitioners identified limited QR awareness as affecting QR documentation. However, its effect is dependent on the project context and role. Communication gaps among ASD team members can create confusion regarding QRs. Missing and outdated QR documentation leads to incurring technical debt, a lack of common understanding of QRs, and incorrect implementations.

We used the knowledge gained from answering RQ1–RQ3, as well as QR documentation practices identified in our prior works with the three cases and review of literature (Behutiye et al., 2020a, 2020c), to answer RQ4 by proposing a model to support optimal QR documentation in ASD. Our prior work with the three cases (Behutiye et al., 2020c), is conducted within the Q-Rapids project and is based on interview data collected during March 2018 and May 2018. The model conceptualizes the factors affecting QR documentation and provides strategies to mitigate issues arising due to these factors. For instance, it shows how time constraints may lead to the under-specification of QRs and that we can use lightweight artifacts, such as a 'given-when-then' template, to document QRs, thus mitigating the challenge of underspecifying QRs due to time constraints.

This study differs from our prior work (Behutiye et al., 2020b) as follows. We include a new contribution, which is the model. We updated the introduction to motivate the need for our study (e.g., showing the research gap in ASD documentation models and guidelines) and explain the delta between this study and our previous work. We updated the related work section by improving the discussion on related topics and showing how the existing work does not address what our work intends to do. In the research methods section, we include a more in-depth description of the cases under investigation and provide a detailed example of the thematic synthesis used in the data analysis. The updates in the results section include rephrasing to clarify the findings and additional quotations providing examples of our findings. Furthermore, a new section is included to present the model. In the discussion, we address the implications of the model and our contribution and update the threats to validity. We also updated the conclusions section to reflect the new contribution.

The remainder of the paper is organized as follows. Section 2 presents the related work. Section 3 describes the research method applied in the study. Section 4 presents the results of the study. In Section 5, we present a model for optimal QR documentation and our plans for evaluating the model. In Section 6, we discuss our findings, the study's implications, and threats to validity. Finally, in Section 7, we conclude the paper.

## 2. RELATED WORK

The existing ASD literature does not explicitly study practitioners' perceptions of the importance of documenting QRs. However, there are studies that focus on understanding documentation practices (Hoda et al., 2012; Stettina et al., 2012; Stettina and Heijstek, 2011; Voigt et al., 2016); and those that focus on exploring QR documentation in ASD (Amorndettawin and Senivongse, 2019; Barbosa and Sampaio, 2015; Behutiye et al., 2020c, 2017; Ho et al., 2006) and understanding practitioners' perceptions of QRs (Kopczyńska et al., 2020). In what follows, we discuss the related work on documentation in ASD and QR documentation in ASD.

## 2.1 Documentation in ASD

ASD advocates minimal documentation practices, as highlighted in one of its four core values, "working software over comprehensive documentation" (Beck et al., 2001). Minimal documentation enables the quick delivery of working software and early returns on investments. However, it is open to misinterpretation. For instance, the focus on minimal documentation is misinterpreted as documentation being unnecessary (Dingsøyr et al., 2012) or as 'just enough' documentation, although it is unclear what 'just enough' documentation entails (Hoda et al., 2012). Such interpretations may be detrimental to software, especially when considering QRs. This is due to the elusive nature of QRs, as they are difficult to define and measure (Paech and Kerlow, 2004).

Regarding documentation in ASD, Hoda et al. (2012) interviewed practitioners from 23 software organizations in India and New Zealand to examine documentation strategies applied in ASD. They found that practitioners use electronic back-ups of paper artefacts, document change decisions made by customers, document business terminologies in project dictionaries to enhance requirement elicitation, and approach collaboration with non-agile teams with traditional documentation. Voigt et al. (2016) studied documentation practices in ASD by employing a theoretical model of information and documentation. The authors found that satisfaction with information searches is correlated with the level of documentation for most types of information. They also revealed that documentation on architecture and design models was insufficient and recommended the development of more methods and tools to support agile documentation. Similarly, Stettina et al. (2012) explored documentation practices and the effect of formalism in ASD in an experiment with students. They found that iterative documentation practices resulted in more detailed textual information. They also found that students perceived writing documentation as an intrusive task and assigned it to less-skilled team members. Although these three studies contribute to the body of knowledge on documentation practices in ASD, none of them examine practitioners' perceptions of documenting QRs in ASD. They did, however, identify research gaps, including the need for more strategies supporting agile documentation, which we aim to address in our paper by proposing a model to support optimal QR documentation in ASD.

A different line of work focuses on understanding practitioners' perceptions of documentation in ASD but does not address QRs (Stettina and Heijstek, 2011). Stettina and Heijstek (2011) investigated practitioners' perceptions of documentation in ASD in a survey study. Their findings reveal that more than half of the 79 respondents in their study considered documentation to be important. They also found that ASD teams adopt collaboration tools in their work (e.g., issue trackers and wikis) to support documentation.

## 2.2 QR documentation in ASD

QRs impact both software quality and cost, which, in turn, affect the success of software projects (Behutiye et al., 2020a; Mendes et al., 2016). Studies reveal that in ASD, practitioners rely on tacit knowledge and tend to avoid documenting QRs (Behutiye et al., 2017; Mohagheghi and Aparicio, 2017). Some practitioners may not document QRs even if they consider them important (Robiolo et al., 2019). These findings prompt the need for understanding practitioners' perceptions of documenting QRs in ASD.

In the ASD literature, there are a few works that explore QR documentation practices (Amorndettawin and Senivongse, 2019; Barbosa and Sampaio, 2015; Behutiye et al., 2020c, 2017; Ho et al., 2006). Among these studies, (Amorndettawin and Senivongse, 2019; Behutiye et al., 2020c, 2017) focus on investigating documentation practices of QRs without differentiating the QR type. Studies by (Barbosa and Sampaio, 2015; Ho et al., 2006) focus on supporting specification of QRs such as performance and security specifications in ASD.

Our previous work (Behutiye et al., 2017) investigated QR documentation practices and challenges in ASD and proposed guidelines for documenting QRs. We found that ASD teams apply artefacts (e.g., user stories, epics, and acceptance criteria), wikis, and backlogs to document QRs and focus on face-to-face communication in smaller teams. Lack of traceability, missing lower-level detail information regarding QRs, and difficulty in documenting internally generated QRs were identified as challenges of documenting QRs in ASD. Our recent work (Behutiye et al., 2020c) explored QR documentation practices in ASD in-depth. We found that QR documentation practices are dependent on the needs of project contexts and affected by the experience of practitioners. We found that companies applied different tools (e.g., JIRA, DOORS, Focal Point), artifacts (e.g., epics, stories, tasks, and prototypes) and practices (e.g. documenting QR decisions, applying guidelines) to document QRs. Practitioners identified the level of abstraction, the traceability of QRs, optimal detail of information of QRs, and verification and validation as important aspects to consider when documenting QRs in requirement management repositories. These works did not examine how factors, such as communication gaps, QR awareness, and time constraints, affect QR documentation in ASD, which we address in this paper.

There are documentation guides and models focused on supporting the documentation of specific QRs. Amorndettawin and Senivongse (2019) introduced a non-functional requirements pattern template to enhance the identification of QRs in ASD by focusing on security and fault tolerance requirements. The authors evaluated the pattern with ten practitioners on a scrum team. They found that the templates helped practitioners write requirements more quickly and comprehensively compared to when a template was not used. Barbosa and Sampaio (2015) proposed a guide to enhance security measures in ASD. The guide includes measures to support the documentation of security QRs by introducing security backlogs, using evil user stories, and providing security training. Ho et al. (2006) proposed an evolutionary model for specifying and validating performance requirements in ASD. The model helps practitioners identify and specify performance QRs incrementally.

Kopczyńska et al (2020) examined practitioners' perceptions of the importance of QRs in ASD without discussing documentation in detail. The authors surveyed 118 ASD practitioners regarding their perceptions of the importance of QRs. They found that about 77% of their respondents perceived defining QRs, i.e., specifying QRs, at least important, and 30% perceived it as critical. Although the study explored practitioners' perceptions of the importance of QRs, it did not explore QR documentation in ASD in detail. For instance, it did not synthesize justifications underlying the importance of defining QRs, which we examine in this paper.

Understanding how practitioners perceive the need for documenting QRs in ASD can help address the research gap in comprehending practitioners' motivation behind their decisions of documenting QRs in ASD. It also provides insight into factors that may affect QR documentation in ASD and is helpful in deriving models and guidelines to support QR documentation in ASD. In our study, we examine practitioners' perceptions regarding the importance of documenting QRs and propose a model to support optimal QR documentation in ASD.

## 3. Research method

### 3.1 Cases

We conducted a multiple case study of three cases. The companies providing the cases vary in terms of size, geographical location, and product domain. Table 1 provides a summary of the cases.

Case A is a medium-sized software development company applying tailored ASD (i.e., it holds daily stand-up meetings, works in sprints, conducts sprint planning and retrospective meetings, and works in 6-month release cycles). It develops a modeling tool and focuses on developing solutions for desktop and Web applications. QR documentation practices of the case differ depending on the project context and the QR. The case emphasizes minimal documentation practices and relies on face-face communication. At times QRs can be known clearly from the outset, and they can be easily communicated in white-board meetings or be specified in word documents. Developers can document QR decisions in word documents as user stories or without following user story formats. The case applies iterative prototypes to document QRs such as performance and usability which may be difficult to specify early on. Case A also utilizes the Mantis tool for documenting QRs and quality issues reported by customers. Project managers and developers also use the Redmine tool to document QRs related to maintenance issues.

Case B is a medium-sized company operating in the telecommunications and embedded systems development domain. It uses Scrum methodology and applies guidelines and the agile-playbook approach to support the documentation and management of QRs. The guideline provides information on QR types (e.g., security, usability) and suggestions on how to document the QRs. It also uses various tools, such as JIRA, to document QRs. QRs are documented in JIRA at different levels of abstraction as epic, story and task. For instance, documenting a QR at epic level involves describing the QR, the verification method, and the Definition of Done (DoD) of the QR, which explain conditions for accepting the epic as completed. The case applies 'Given/when/then' template to specify QRs in DoDs of stories and tasks. There are also separate organizations that are responsible for specifying and managing specific QRs, such as security. When specifying and documenting QRs such as security, Case B also applies additional guidelines that take into account security standards and certifications.

Case C is a large, multinational telecommunications company. It adopts large-scale distributed ASD and applies varying practices based on organizational levels and types of QRs. For instance, at lower levels, teams may use Post-it notes, whereas different tools, such as Focal Point, DOORS, Accept 360, and JIRA, may be used at higher levels to document and manage QRs. Case C documents QRs in JIRA following different levels of abstraction as feature items, system entities, entity items, competence area items, epics, tasks, sub-tasks. It applies DoD to document QRs of tasks too. In the case, QRs can also be documented as items of improvement backlog, which is a backlog used to document improvement ideas that are uncovered during the software development process. The case also uses separate organizations to handle the documentation and management of some QRs, such as security and performance.

*Table 1. Summary of cases*

| Case | Product domain | Company size in number of employees | Software development method | QR documentation practices and tools | Quality drivers and related documentation practice | ASD adoption (years) |
|---|---|---|---|---|---|---|
| A | Modelling tool | Over 900 | Based on ASD principles | Minimal documentation emphasis, relies on face-face communication, Mantis to document quality issues, document QR decisions in word documents | Performance and usability Iterative prototypes to document QRs | 15 |
| B | Telecommunication, Embedded systems | Over 600 | Scrum | Guidelines, agile playbook to document and manage QRs, JIRA, QRs documented as epic, stories and tasks, applies DoDs | Security, performance Separate organizations to document and manage the QRs | 14 |
| C | Telecommunication | Over 100,000 | Large scale distributed ASD | Varying practices, tools such as Focal Point, DOORS, Accept 360, JIRA, QRs documented as features, system entities, competence area items, epic, task, sub task, documented as DoDs of tasks, QRs documented as improvement backlog items | Security, performance Separate organizations to document and manage the QRs | 12 |

### 3.2 Study design and data collection

We designed a protocol for the multiple case study based on Runeson and Höst's (2009) case study guidelines and conducted semi-structured interviews with 12 ASD practitioners employed by the three cases. Prior to the interviews, we communicated our research objective and applied the key informant technique (Marshall, 1996) to suggest interview participants based on their roles, such as project managers and requirements specification engineers, to the representatives of the cases. We used the key informant technique, as it offers collecting high-quality and valuable evidence from experts on a specific topic (Marshall, 1996). The representatives of the cases reviewed our suggestions, identified the relevant participants, and helped arrange meetings to interview the participants. The interviews were conducted between September and October 2019. Each interview was audio recorded and lasted between 25 and 35 minutes. Table 2 shows data regarding the interview participants' backgrounds. The interview participants' backgrounds varied in terms of their experience in software engineering and ASD. The participants' median experience in software engineering was 16.5 years, whereas the participants' median experience in ASD was 10 years. Among the participants, the minimum software engineering experience was 12 years, whereas the maximum was 24. Regarding ASD experience, two years was the minimum, whereas 15 was the maximum. Appendix A provides the interview script.

*Table 2. Summary of the interview participants' backgrounds*

| ID | Case | Role | Software engineering experience (Years) | ASD experience (years) |
|---|---|---|---|---|
| P1 | A | Project manager | 21 | 11 |
| P2 | A | Project manager | 12 | 3 |
| P3 | A | Executive manager | 31 | 14 |
| P4 | A | Software architect | 12 | 12 |
| P5 | B | DevOps tech lead | 17 | 15 |
| P6 | B | Process coach | 16 | 7 |
| P7 | B | Build manager | 14 | 4 |
| P8 | B | Project manager | 10 | 10 |
| P9 | C | Requirements specification engineer | 24 | 10 |
| P10 | C | Software architect team lead | 20 | 10 |
| P11 | C | Requirements specification engineer | 24 | 7 |
| P12 | C | Product architect lead | 15 | 2 |

### 3.3 Data analysis

We applied thematic analysis (Cruzes and Dybå, 2011) to analyze the collected data. First, the audio recordings of interviews were transcribed through a professional service. Then, the first author read the interview transcriptions and labeled excerpts describing concepts that answer our research questions with codes in NVivo, which is a qualitative analysis tool. For instance, we coded excerpts describing the importance of QR documentation in ASD in each of the cases as 'significance of documentation of QRs in ASD.' We followed similar approaches to label concepts regarding the consequences of missing or outdated QR documentation and factors affecting QR documentation in ASD (e.g., QR awareness and the effects of time constraints). Next, within each of the cases, we compared the concepts gathered under each of the codes with each other, and those that were related or recurring were refined and grouped into a theme. There were also non-recurring concepts. This step resulted in themes and non-recurring labels answering our research questions within each of the cases. Then, we compared themes and labels identified from each of the cases with each other and grouped closely related and similar themes, thereby refining them into final lists of bigger themes. Figure 1 illustrates the theme "documentation left as the last thing to do," which we derived by applying thematic synthesis to excerpts coded as the effect of time constraints on documentation.

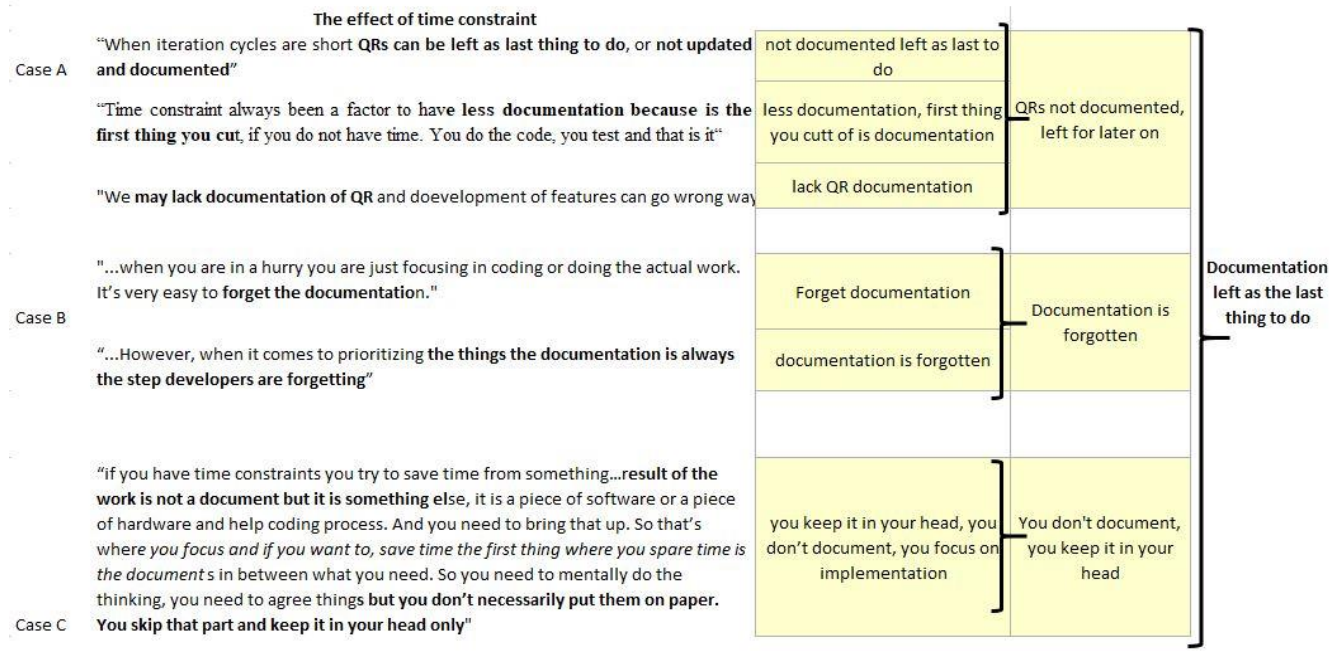

*Figure 1. Thematic synthesis example.*

## 4. RESULTS

In this section, we provide answers to our research questions. Section 4.1 presents the practitioners' perceptions of documenting QRs in ASD. Section 4.2 presents the practitioners' understanding of the effect of time constraints, QR awareness, and communication gaps regarding QR documentation. In Section 4.3, we address the consequences of missing and outdated QR documentation.

### 4.1 ASD practitioners' perceptions of documenting QRs (RQ1)

All participants in the study agreed that in ASD, it is important to document QRs. However, they noted that the QR type and project context determine the level of documentation needed. For instance, the executive manager from case A explained, *"Yes, quality requirements are important to document. The quality requirements are inherently part of the requirements for projects. In some projects, if you have safety aspects, for example, if you make a project on autonomous cars, quality is utterly important. Or, if you have just the utility that you use just twice a year, quality is not that much important."* The practitioners identified documenting QRs as important to ensure quality, clarify QRs, establish process conformance, and assist decision-making on QRs, as well as because QRs influence the implementation of other features. Ensuring the quality and clarity of QRs are the top two recurring themes we found in all three cases. Table 3 summarizes the practitioners' justifications for documenting QRs in ASD. In the table, an X indicates that responses from at least one or more participants of a case are mapped to the theme on the corresponding row.

*Table 3. Practitioners' justifications of the importance of documenting QRs in ASD*

| Justification for documenting QRs in ASD | Case A | Case B | Case C |
|---|---|---|---|

| Ensuring quality | X | X | X |
|---|---|---|---|
| Clarity of QRs | X | X | X |
| Process conformance | X | X | - |
| Help in decision making | - | X | X |
| Influence implementation of other features | X | - | X |

### 4.1.1 Ensuring quality

Practitioners in the three cases stated that documenting QRs is important to ensure quality. They explained that documenting QRs helps establish the required quality targets and the acceptable level of quality for QRs. For instance, the DevOps tech lead from case B explained how documenting QRs supports ensuring quality*:*

> *I think this kind of correct syntax for providing some acceptance testing and stuff like this, to get some gatekeeping levels for what is acceptable quality, and then those kinds of requirements link together as a cluster, having a whole understanding of what quality requirements go together and why. So, these kinds of tools like Jira support that kind of thing that you can use the given-when-then format to create acceptance testing and acceptance requirements that really show the acceptable quality and acceptable test cases.*

Likewise, a software architect from case A explained that documenting QRs helps ensure quality by supporting verification: "*It is important to verify the way we can answer quality issues, what are the tasks, the concrete tasks to do to answer the quality issue, I think. And to do that, documentation of QRs is required.*" The requirements specification engineer from case C described how documenting QRs helps define the desired quality targets and guide related work: *"But for us, who are working with these systems, for example, the capacity issues are pretty well documented. The key quality targets are defined, and they guide our work. So, I think that those are good to document."*

### 4.1.2 Clarity of QRs

Practitioners in all cases identified documenting QRs as important to clarify the meaning and scope of QRs. This is important since requirements often change, and the QRs' definitions may evolve over time. A project manager from case A explained how documentation supports the clarity of QRs: *"Well, I think that it is important to document them. That always makes things clearer, and of course, it will also help the same person who wrote that description."* Similarly, the build manager from case B explained, *"I think it has beneficial value. When you have quality requirements defined and documented properly, they become more clear.*" We also observed that QRs may not always be understandable to all stakeholders and that documenting them may help in clarifying them. The software architect team lead in case C stated*, "But, unless those basic things like troubleshooting, software updates, resiliency, robustness, if they are not in place, no customer or product manager is going to upfront say that okay, those are needed. But, they are assumed to be there, so they need to be there properly framed. And when developing the next releases of the products, unless those are documented well in the original development of the product, they are easily forgotten."*

### 4.1.3 Process conformance

Practitioners from cases A and B identified documenting QRs as useful in helping to ensure process conformance. Their responses reveal how documenting QRs can help in monitoring and ensuring that the practices and activities are followed as recommended in the cases. For instance, a project manager from case A explained, *"The idea is to always follow the same process and be sure that part of the development was not or was taken into account by the project manager, for example. Quality is something that takes a lot of time, and the fact that we formalize how you have to answer a quality issue forces you to take the time to resolve the issue."* Similarly, the DevOps tech lead from case B explained how documentation helps in conforming to the specification process: *"I think that is the important thing, that people follow the same format and understand which of those requirements have been clustered together and why."*

### 4.1.4 Help in decision-making

Two interviewees from cases B and C stated that documenting QRs may enhance decision-making during the elicitation, implementation, and validation process. Properly defining and documenting QRs ensures what has been agreed upon regarding the QRs among team members and thus facilitates decision making. According to these interviewees, unless QRs are documented, it is difficult to know what has been agreed upon previously regarding the QRs, as they can easily be forgotten. For instance, the build manager from case B explained, "*If quality requirements are defined properly, it helps the decision making.*" Similarly, the requirements specification engineer from case C explained,

> *When we have agreed on what we are doing, of course there are some non-functional parts that also need to be agreed on; for example, how many users we will have. . . . That, I think, can be thought of as a non-functional or quality requirement. So, if we are, first we are agreeing on something, and then it's written down what has been agreed on, it is then this written specification or written requirement. So, if we are agreeing on something, but we don't write it down, who knows what we agreed on. So, in that sense, documenting everything that has been agreed on is important.*

### 4.1.5 QRs influence on the implementation of other features

Two practitioners from cases A and C explained that QRs should be documented since they may affect the implementation of other features. For instance, the project manager from case A explained, *"When you get feedback from the testing team, the client, users, or the tool itself, feedback needs to be related to some quality requirement. In our context, the feedback is so large and can impact many features or functionalities; that is why we need to keep track by documenting the quality requirements."*

## 4.2 ASD Practitioners' perception of factors that may influence documenting QRs in ASD (RQ2)

We collected practitioners' opinions on whether time constraints and iteration cycles in ASD, QR awareness, and communication gaps on QRs among development team members may or may not influence documenting QRs in ASD. Table 4 summarizes themes identified under the factors that may influence QR documentation in ASD. In the table, an X shows that a response from at least one participant or more participants of a case, are mapped to the theme on the corresponding row.

*Table 4. Summary of factors influencing QR documentation in ASD*

| Factors affecting QR documentation in ASD | Themes | Case A | Case B | Case C |
|---|---|---|---|---|
| The effect of time constraints and short iteration cycles on documenting QRs in ASD | Documentation of QRs left as the last thing to do | X | X | X |
| | QRs not defined and specified well, and led to rework and additional iterations | X | X | - |
| | Depends on sprint duration and project | - | X | - |
| | Difficult to document when there are feature dependencies | - | - | X |
| | Time constraints and short iterations do not affect documentation of QRs | - | - | X |
| The effect of QR awareness on documenting QRs in ASD | QR awareness of practitioners and customers affects QR documentation | X | X | X |
| | Depends on the role | X | X | - |
| | Depends on the project context | - | X | - |
| | Depends on the QR type | - | - | X |
| Opinion on the effect of communication gap on documenting QRs in ASD | Communication gap affects documentation | X | X | X |
| | Communication gap will not arise if QRs are specified and documented in the early stages | X | X | - |
| | Can create confusion | - | X | - |

### 4.2.1 The effect of time constraints and short iteration cycles on documenting QRs in ASD

Except for one interviewee, all participants stated that time constraints and short iteration cycles affect QR documentation in ASD.

a) *Documentation of QRs left as the last thing to do:* Six *interviewees* from the three cases reported that time constraints and short iteration cycles in ASD affect QR documentation. When facing time constraints, QR documentation becomes less important, and developers focus on implementing the software. When QRs are postponed and 'the last thing to do,' they are not updated and documented and are easily forgotten. For example, the executive manager from case A explained, *"Time constraint has always been a factor in having less documentation because it is the first thing you cut if you do not have time. You do the code, you test, and that is it."* Likewise, the process coach from case B explained, "*When you are in a hurry, you are just focusing on coding or doing the actual work. It's very easy to forget the documentation.*"

b) *QRs were not defined and specified well and led to rework and additional iterations:* Three interviewees from cases A and B stated that time constraints may result in poorly defined QR documents. They also noted that when QRs are not specified and documented properly, feature implementations may not work as intended and require additional iterations to improve the QRs and documentation. For instance, the software architect from case A explained, "*It means that if we didn't solve, invest some time to address the quality issue, we will have problem in the next implementation or in one year, in two years, on this topic. Yes, it's costly in terms of time. If we neglect the quality on some aspect, we know that we will have to invest a lot of time the next time we have to redesign, rework the features.*" Similarly, the build manager from case B noted, "I*n my experience, I have seen that this causes the quality of documentation to be not good at first. You need to make multiple iterations*."

c) *Depends on the sprint duration and the project:* The DevOps tech lead from case B reported that whether time constraints affect QR documentation depends on the sprint duration (e.g., two weeks, three weeks, or a month) and the project context. He explained that achieving good documentation may be difficult in a short sprint duration (e.g., two weeks) on some projects. In such cases, ASD teams apply additional practices to ensure a good level of QR documentation. For instance, when the sprint duration is short, teams may allocate themes for multiple sprints (e.g., one sprint focuses on implementation and another sprint focuses on QR documentation and other improvements). However, they may include documentation and implementation in one long sprint duration, as shown in his reply: *"There are themes that, in one sprint, not everything is done usually. It is more about features, bug fixes, documentation, another feature sprint, another bug fixing sprint, another documentation sprint. Or, if you have a longer sprint, then you can try to do all of those inside one sprint."*

d) *Difficult to document when there are feature dependencies*: The requirements specification engineer from case C explained that in short iteration cycles, documenting QRs is problematic due to feature dependencies. He noted, *"Of course, there are those independencies between features. That is problematic. If we have new features, it's quite obvious that those will be documented well, and we have good background information already, so those are in better shape. But, the features coming to the system and those independencies could not be foreseen. From that, we have learned that it is problematic."*

e) *Time constraints and short iterations do not affect documentation of QRs*: In case C, the requirements specification engineer argued that time constraints do not affect QR documentation. According to him, when multiple teams work together and share a similar goal, time constraints do not influence QR documentation. He explained, *"I am still a little bit skeptical in that, even if we have short iterations and you get the feedback from other teams or a customer or wherever, you can fix it. But, then when we have a lot of teams doing things to meet the same goal, and those teams' work cannot be isolated so well, I think it does not minimize or decrease the need for documentation."*

4.2.2  The effect of QR awareness on documenting QRs in ASD

We asked the interviewees whether QR awareness affects documentation. All the interviewees pointed out that QR awareness may affect QR documentation. We identified four themes on the effect of QR awareness on QR documentation in ASD.

a) *QR awareness of practitioners and customers affects QR documentation:* This was shared by nine interviewees of the three cases. The interviewees stated that in ASD, stakeholders' knowledge of QRs affects the likelihood of documenting QRs. For instance, a project manager from case A described how practitioners' knowledge of QRs affects documentation, *"When you design or when you think about your quality requirements, it is related to some particular issue, what is the main, the core, the important feature of your tool, what it is part of? If it is security, if it is code quality, and by thinking about or defining the quality requirement, you also think about what it is needed the documentation . . . If you do not think in terms of quality requirements, it will have an* *impact on the documentation."* The software architecture team lead from case C noted that junior developers may not be familiar with QRs and indicated practices that help in identifying and documenting QRs:

> *Unless you have that kind of checklist of the quality areas, the non-functional requirement domains to be documented, and different views on the product behavior, they are not that easy to invent by yourself. Even though you could be inventing them yourself and figuring them out, that easily leads to a very fragmented categorization of those non-functional requirements. So, having a canonical way of looking at the non-functional requirement area helps with the documentation as well.*

However, it is not only ASD practitioners' knowledge of QRs that affects QR documentation. For example, the build manager from case B shared his experience with how inadequate QR knowledge by customers and the lack of QR documentation affected their work:

> *At a previous company I was doing sub-contracting for, the customer had neglected the quality requirements, and we struggled to get the needed information from them because they did not have their own. It affected our work and caused the delivery to be delayed because the quality requirements on their part were neglected. They did not have much knowledge of quality requirements, and combined with the lack of documentation, it affected our work.*

b) *Depends on the role:* A project manager from case A and the process coach from case B noted that whether QR awareness affects QR documentation in ASD depends on the role. For instance, the project manager explained that people in certain roles, such as product owners and project managers, should care about and have good knowledge of QRs. This is because they need to decide whether the quality level of a product is met and ensure product readiness. However, the interviewee argued that some roles, such as a developer, may place more emphasis on development and not care much about QRs and their documentation. The process coach from case B pointed out that poor specifications and QR documentation carried out by managerial roles at higher levels may affect how developers at lower levels document and implement QRs.

c) *Depends on the project context:* According to the DevOps tech lead in case B, whether QR awareness affects QR documentation depends on the project context. He stated that certain projects may value functional requirements over QRs and treat QRs as an afterthought that will be carried out by the quality organization. However, other projects may value and be thoughtful about QRs; thus, these projects emphasize QR documentation.

d) *Depends on the QR type:* According to the requirements specification engineer from case C, it is difficult to specify and document some QRs, such as capacity, in the early phases of development because they may require rework and even hardware changes:

> *We can set some targets that we want to reach; for example, we want to have 500 users at the same time. But then, at some point, when all functional requirements are implemented and tested, there might be some surprises, such as we are not reaching what we set at an earlier point as the capacity target. So, it might be that we are getting only 450 for some reason. At that point, it is very expensive to start over. To meet all those functional requirements, maybe even the hardware needs to be updated to reach the original goal of 500. So, it means that it is not so easy to define these non-functional requirements in early phases.*

### 4.2.3 The effect of communication gaps regarding QRs among team members on documenting QRs in ASD

We asked the interviewees whether communication gaps on QRs among ASD team members affect QR documentation. Except for two of the interviewees, all agreed that communication gaps regarding QRs may affect documenting them in ASD.

a) *Communication gaps affect the documentation of QRs:* Nine interviewees from all cases explained that communication gaps among team members regarding QRs can affect QR documentation. The interviewees described how members' absences, misunderstandings and misinterpretation of QRs, not discussing QRs, and missing QR documentation can create communication gaps. Indeed, these lead to either not documenting QRs or poor QR documentation. However, the interviewees suggested practices that help minimize the communication gaps. For instance, they recommended open discussions of QRs among team members to minimize misunderstandings of QRs and improve QR documentation in ASD. For instance, the build manager from case B explained, *"Yes. I think it also affects that. More open discussion will help in improving the documentation."* Similarly, the software architect from case A explained, *"When we deliver the first release or code, the testing team launches their testing or makes their test, and if we get feedback on quality issues, we discuss the documentation of this quality requirement. If we didn't understand the same thing in this quality requirement documentation, we document it a little bit better or longer or describe it a bit more clearly."*

b) *Communication gaps will not arise if QRs are specified and documented in the early stages:* Two interviewees from cases A and B presumed that there will not be communication gaps regarding QRs among team members if QRs are agreed upon and specified at the beginning of the project. For instance, the project manager from case A explained, *"I cannot think that a misunderstanding of quality requirements can really happen because you set them from the beginning. So, if you decide, for example, to have certain accessibility because your clients have some particular needs, you just state that from the beginning, and you define it."* Documenting QRs in the early stages is assumed to facilitate the communication of QRs, hence preventing QR communication gaps among team members.

c) *Communication gaps can create confusion:* The DevOps tech lead in case B argued that if there are communication gaps among team members regarding QRs, it is likely that team members document QRs in their own way, which can create confusion on the tasks and unnecessary conflicts. He suggested that clear and early specification and documentation of QRs may help prevent such confusion: *"If there's a gap, when the review for that documentation comes, there will be at least three people who say that I don't agree with your thresholds, why haven't we talked about this, why we are aiming for 50 per cent of this, why not 60 per cent? I believe it will create a lot of confusion and conflict, which is unnecessary because, if they talk it through beforehand, before something's written down, then there's no need to fight over it."*

## 4.3 ASD practitioners' perceptions of the consequences of missing and outdated QR documentation (RQ3)

We identified five consequences of missing and outdated QR documentation in ASD. We summarize these consequences in Table 5 and discuss them below. In the table, an X shows that responses from at least one participant or more participants of a case, are mapped to the theme on the corresponding row.

*Table 5. Consequences of missing or outdated QR documentation in ASD*

| Consequence of missing or outdated QR documentation in ASD | Case A | Case B | Case C |
|---|---|---|---|
| Technical debt accumulation | X | X | X |
| Practitioners may not know what the QRs cover and lack the understanding of the current behavior | X | X | - |
| Lack of common understanding of QRs | X | - | X |
| Informal quality management process | X | - | - |
| Incorrect implementation leading to unhappy customer | - | - | X |

a) *Technical debt accumulation:* Four interviewees from all cases explicitly stated that missing and outdated QR documentation in ASD may lead to the accumulation of technical debt. Additionally, others reported system quality and performance degradation (two interviewees from cases B and C), increased development time (three interviewees from all cases), increased maintenance costs (two interviewees from B and C) and rework (one interviewee from case C), which are indicators of technical debt, as the consequences of missing and outdated documentation.

b) *Practitioners may not know what the QRs cover and lack the understanding of the current behavior:* Three interviewees from cases A and B reported that when QRs are not documented, it will be unclear what the QRs cover. For instance, the DevOps tech lead from case B explained, *"For me, the first thing that comes to mind is that I do not know when I am ready; I am not ready when I have done the features, I want to know how it has been used, is it working well? If I do not have any quality metrics and quality requirements to tow these metrics, I really don't know that. I have the functionality. I have no understanding of the current behavior of the usage model, or is it actually useful or anything like that."* Moreover, in such scenarios, practitioners may become frustrated and demotivated as they spend additional time revisiting old features.

c) *Lack of common understanding of QRs:* Three interviewees from cases A and C reported that missing QR documentation may result in a lack of shared understanding of QRs and create confusion among practitioners. Multiple interpretations of QRs can further lead to friction among team members. However, practitioners suggested practices to mitigate this problem. For instance, a project manager from case A suggested that ensuring the right level of QR documentation may help in avoiding confusion regarding QRs that results from missing and outdated QRs.

d) *Informal quality management process*: A project manager from case A noted that missing and outdated QR documentation may lead to informal quality management processes. Moreover, she explained that it may indicate a lack of quality focus in the process: "*If the quality requirements are not documented, I think that it's proof that the quality process is not at the center of development.*"

e) *Incorrect implementations leading to unhappy customers:* The requirements specification engineer from case C revealed that not documenting QRs may result in incorrect implementation and unhappy customers: "*The worst-case scenario is if those basic quality requirements are not documented. Then, they are not implemented into the product, and the lack of some basic capability, recovery capability, and so on is only found until after the product has been shipped to the field, and we have an unhappy customer.*"

## 5. MODEL TO SUPPORT OPTIMAL QR DOCUMENTATION IN ASD

In this section, we present an initial version of a model for supporting optimal QR documentation in ASD and our plan for evaluating the model. Section 5.1 introduces our model and Section 5.2 discusses the evaluation plan.

### 5.1 Introducing a model for optimal QR documentation in ASD

We answer RQ4 by introducing an initial version of a model that has an ultimate goal of providing support for optimal QR documentation in ASD. Models provide means of representing knowledge gained from observed reality (Shaw, 2002). We built the model by considering the gap in models and guidelines that support the documentation and management of QRs in ASD. In our review of the management of QRs in ASD (Behutiye et al., 2020a), we noticed that existing strategies for the documentation and management of QRs overlooked the effect of some factors (e.g., time constraints, QR knowledge, and communication gaps) on documenting QRs. These existing strategies often addressed the limitations of ASD artifacts in specifying and documenting QRs (e.g., the limitations of user stories in specifying QRs or only addressing specific types of QRs, such as security, performance, and usability). The scope of our model covers the factors: time constraints, limited QR awareness and communication gaps among team members. It is a first step in our work and targets supporting optimal QR documentation in ASD as well as improved understanding regarding QR documentation in ASD. It provides insights into how time constraints, QR awareness, and communication gaps among team members, affect QR documentation in ASD. It also explains the issues in QR documentation that may arise from these factors and present potential list of corresponding mitigation strategies for addressing them. Additionally, it explains the dependency of the aforementioned factors from project and role perspectives and reveals the consequences of missing and outdated QR documentation.

We built the model based on the findings of RQ1–RQ3 and practices used to document and manage QRs in ASD, which we identified in our review of the literature on QR management in ASD (Behutiye et al., 2020a) and an investigation of QR documentation practices in the three cases of our study that we conducted previously (Behutiye et al., 2020c). We used the practices from these studies to propose mitigation strategies for issues arising from factors affecting QR documentation. Figure 2 shows an illustration of the model. In the model, the first layer describes three factors that affect QR documentation. The second layer characterizes the corresponding effects of these factors. For instance, it maps the effect of limited QR awareness to 'QRs neglected and not documented.' The third layer, project dependency, describes whether the factor influencing QR documentation depends on the project characteristic (e.g., sprint duration used in the development process, QR documentation and management practices, product domain, and resource availability). The prevalence in certain roles layer helps explain whether a factor influencing QR documentation is more common to a particular role (e.g., the product owner, project manager, customer, or ASD team, including developers, the scrum master, the product owner, testers, and software architects). The next layer depicts the consequences of underspecified and outdated QR documentation. Finally, in the mitigation strategies layer, we present different strategies to mitigate the effects (the issues) that may arise due to the identified factors. In the following paragraphs, we explain each layer in detail.

1. *Factors affecting QR documentation*: time constraints, limited QR awareness and communication gaps among team members. Time constraints refer to the limited time available due to the short iteration cycles of ASD. Limited QR awareness refers to the lack of knowledge regarding QRs. Communication gaps among team members refer to miscommunication regarding QRs that can occur among ASD team members.
2. *Effects on QR documentation:* explains the effects that may occur due to the aforementioned factors. For example, when considering time constraints, we observe that it may result in postponing QRs, not documenting them, poorly specifying them, and omitting them from the implementation. Additionally, time constraints may lead teams to incur additional iterations for documentation work. Limited QR awareness among practitioners may lead to neglecting and not documenting QRs. Junior developers may not have enough knowledge of QRs and may find documenting and implementing QRs difficult. Additionally, limited QR awareness can lead to not knowing about required system behavior (e.g., security or performance), making documenting and implementing QRs more challenging. When communication gaps exist among team members, QRs become unclear, and there is a greater likelihood of friction or conflict among team members due to confusion regarding QRs.
3. *Project dependency*:

    a) When considering time constraints, we observe that its effect on QR documentation can depend on project characteristics (e.g., sprint duration used in the development process and resource availability for QR documentation). For instance, in a project with a two-week sprint duration, ASD teams may find it difficult to complete an adequate amount of QR documentation. However, on projects with a longer sprint duration (e.g., longer than two weeks), the effect of time constraints on documenting QRs is minimal since the long sprint duration offers enough time for documentation. Additionally, projects with insufficient resources for documenting and managing QRs underspecify and fail to document QRs when there is a time constraint. Projects that have sufficient resource allocation and clear goals regarding QRs and their documentation may not face similar problems.
    b) Whether QR awareness affects documentation depends on project characteristics. For instance, some projects can establish process for documenting and managing QRs. In these projects, the likelihood that issues arise from limited QR awareness is minimal since there is an established process for documenting and managing QRs. However, if projects do not have a process for documenting and managing QRs, limited knowledge of QRs among the ASD stakeholders can create challenges in documenting QRs. Additionally, in projects where hardware is involved (e.g., in the embedded domain), identifying and specifying QRs early on is difficult. This is because QRs evolve and may require hardware changes.
    c) Whether a communication gap among team members affects QR documentation depends on project characteristics such as practices for QR documentation and management. For instance, projects that specify and document QRs clearly and early in the development process and have established practices to resolve communication gaps (e.g., openly discussing unclear QRs and resolving QR issues) are less likely to be affected than projects that do not practice early specification or openly discuss QR issues. In the case of large-scale distributed ASD setting, if QRs are not clearly specified and communicated among team members, managing them becomes challenging.
4. *Prevalence in certain roles:* We examine whether each factor that affects documenting QRs is more particular to specific roles or affects all stakeholders to the same extent.

    a) We observed that time constraints affecting QR documentation are not more prevalent in some roles than in others. Instead, the effect of time constraints is more dependent on how the whole ASD team functions. The likelihood that time constraints affect QR documentation is minimal in collaborative ASD teams with sufficient resources and clear responsibilities regarding QRs. However, when the responsibilities for documenting and managing QRs are unclear or ambiguities exist in the QRs, time constraints tend to affect QR documentation (e.g., QRs are neglected and are not specified or documented) (Cajander et al., 2013).

b) Limited QR awareness affecting QR documentation is more prevalent in some roles, such as junior developers, since they may lack knowledge of QRs. Additionally, it is highly important that project managers and product owners are well oriented and have adequate knowledge of QRs since they specify the requirements and decide on the quality readiness of products. Customers' knowledge of QRs also affects the documentation and implementation of QRs. Customers who do not understand the significance of QRs may not allocate enough funding in the budget to implement and manage QRs (Camacho et al., 2016). On projects with established processes for documenting and managing QRs, knowledge of QRs in some roles, such as developers, may not be as important since there are clear responsibilities and procedures for documenting and managing QRs. However, when there are no established processes, it is important that developers have good knowledge of QRs.
c) How communication gaps among team members affect QR documentation is dependent on how the whole ASD team functions. ASD teams with established practices and clearly allocated roles for documenting and managing QRs are less likely to face QR documentation issues due to communication gaps among team members.

5. *Consequence of underspecified and outdated QRs:* QRs can become underspecified and outdated due to various factors, such as time constraints and limited QR awareness (Behutiye et al., 2020b). Moreover, there can be technical debt accumulation, performance and system degradation, incorrect or unsuitable implementation, and rework. When ASD team members lack a common understanding of QRs or practitioners do not have a sufficient understanding of the system's behavior, customers can become unhappy.

6. *Mitigation strategies:* We describe strategies to mitigate issues caused by time constraints, limited QR awareness, and communication gaps among team members regarding QR documentation.
   a. Mitigating the effect of time constraints
      i. Lightweight artifacts to document QRs: teams can adopt lightweight artifacts that are compatible with the short iteration cycles of ASD.
         1. Teams can adopt '*given-when-then*' templates to specify QRs as acceptance criteria (e.g., acceptance criteria of a reliability QR in a mobile travel app can be specified as, **Given** <the app is launched and in the homepage> **When <** the user clicks 'My trips' button> **then**< the available trips page should load under 3 seconds>.
         2. ASD teams can also document QRs by including them as acceptance criteria for user stories. For instance, a security QR for checking the SSL certificate of the log-in webpage of an e-commerce site can be written as acceptance criteria of a user story as follows:
         **As a** <user >, **I want to** <log in to the webpage safely>,
         **So that** < I can make online purchases >
         **Acceptance criteria**: Perform SSL (Secure Sockets Layer) test for the page (Behutiye et al., 2020c).
      ii. Teams can tailor their practices to achieve optimal QR documentation in their process.
         1. Scrum teams may adopt theme-oriented sprints (e.g. a sprint for implementation and a sprint for documentation). In a documentation sprint, the teams handle documentation tasks and make improvements regarding QRs.
         2. Alternatively, teams may allocate separate organizations that are responsible for documenting and managing QRs. This is more relevant in the embedded systems and telecommunications domain, as well as in product domains that are required to meet specific regulations and standard certifications. For instance, companies adopt separate organizations and dedicated teams responsible for handling the security and performance QRs of their products. These organizations produce and manage related QR documentation (Alnatheer et al., 2014; Behutiye et al., 2020c, 2020a).
      iii. Allocate sufficient resources and assign clear responsibilities: assigning sufficient resources and clear responsibility to document and handle QRs can help mitigate issues that may occur

due to time constraints. Moreover, when the ASD team works collaboratively, there are adequate resources and clear responsibilities regarding QRs. Everyone works to achieve a common goal, and thus, ASD teams can minimize documentation issues that may occur due to time constraints.

b. Mitigating the effect of limited QR awareness
   i. Raise QR awareness among members of the ASD team:
      1. Invest in QR training: ASD teams should work on developing their knowledge of documenting and managing QRs. Management should invest in training regarding QRs to familiarize ASD team members with relevant QRs and practices to document and handle them. It is highly important to ensure that some roles, such as project managers and product owners, have good knowledge of QRs. Training on QRs is particularly beneficial to junior developers who may not be familiar with QRs and practices for documenting and managing them in the company.
      2. Utilize a QR checklist: teams can adopt a checklist for relevant QRs, including how to specify, document, and manage them within the company. Such a checklist will help developers and other ASD team members become acquainted with QRs and the recommended practices involved in documenting and managing them. Moreover, since some practitioners (e.g., product owners) are prone to forgetting QRs (Behutiye et al., 2020a; Nawrocki et al., 2014), a QR checklist is valuable.
   ii. Guiding customers on QRs: limited QR awareness among customers affects QR documentation (Behutiye et al., 2020a, 2020b; Sillitti and Succi, 2005). For instance, customers may not be aware of all the relevant QRs needed for their product in the early phases. In such cases, practitioners (e.g., product owners and business analysts) should help and guide customers in identifying and specifying QRs.

c. Mitigating the effect of communication gaps among team members
   i. Checklist of QRs: a checklist of QRs, along with detailed instructions for documenting and managing them, can help minimize issues arising from communication gaps among ASD team members (Behutiye et al., 2020c).
   ii. Documenting QR quality targets and decisions: another means to mitigate QR documentation issues due to communication gaps is to practice documenting quality targets and decisions regarding QR tasks. Teams document either their decisions on QRs, or the quality target of the QR. This is especially useful for small teams that work collaboratively and rely on face-to-face communication and white-board discussions when documenting and managing QRs (Behutiye et al., 2020c). For instance, QR decision regarding security QR of software can be documented as follows together with the justification of the QR decision, decision Id and the affected software component.

*Table 6 Documenting QR decisions*

| Field | Description |
| --- | --- |
| Decision ID | 11 (Identifies the decision with unique ID ) |
| Decision summary | Run penetration test |
| Justification | To detect and identify potential vulnerabilities |
| Affected component | Software component Z |

In some cases, teams can document quality target of the QR. For instance, quality target for performance QR, a mobile app can be documented as, "The response time for starting mobile app X should be under 100 millisecond".

   iii. Early on, clear, precise, and testable QR specification: teams should document QRs early in the development process, along with the functional requirements. Specifications of QRs should be brief, clear, meaningful, and testable. This approach helps minimize the likelihood of communication gaps occurring among team members and affecting the documentation and implementation of QRs. For instance, we can specify testability QR at story level as follows.

*Table 7 Clear, precise and testable QR specification*

| Field | Description |
|---|---|
| Issue type | Story |
| Story | Add test cases to reach test coverage of 90%. |
| DoD | With additional test cases the test coverage should reach 90%. |

    iv. Process to document and manage QRs: Establishing a process to document and manage QRs mitigates challenges that may arise due to communication gaps. ASD teams can adopt a process to guide their members on documenting and managing QRs.

Allocation of enough time, good QR awareness among stakeholders and good communication among team members support optimal QR documentation.

In our model and the context of this study, QR documentation describes activities related to defining QRs, recording QR tasks, QR targets and QR decisions during the software development and maintenance processes. We consider QR specification as a process for defining QRs. QR specification can be done initially, and during the software development process, since requirements can evolve in ASD. The 'QR targets' refer to quality goals of the QR that are set by agile teams. The 'QR decisions' referred to decisions regarding QR tasks that agile teams agreed up on and document. QR targets and decisions can be documented at the outset or during the software development and maintenance processes too.

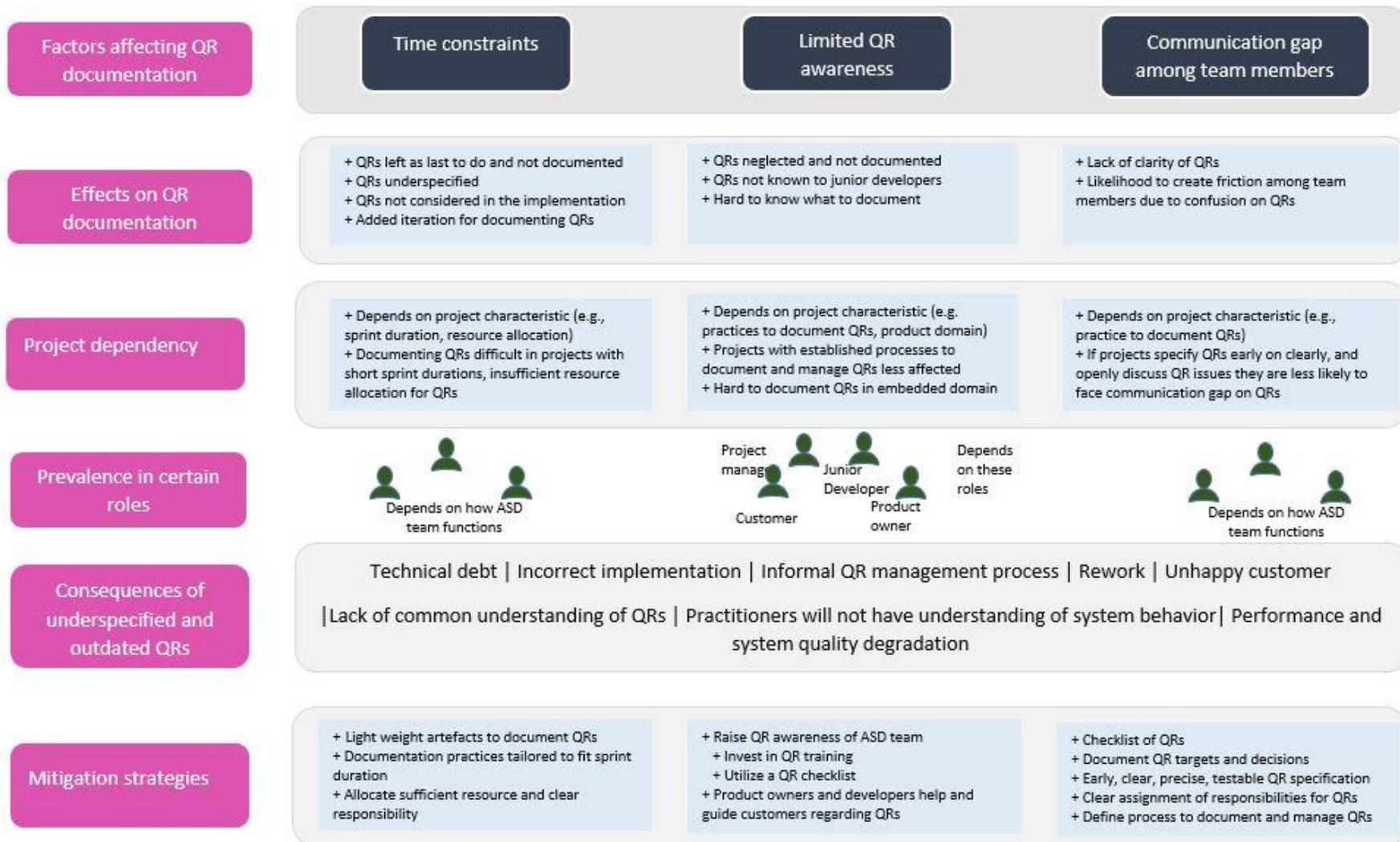

*Figure 2. Model to support optimal QR documentation in ASD.*

## 5.2 Evaluation plans

As an immediate evaluation goal, we plan to evaluate the model with ASD practitioners and experts through survey questionnaires. We justify our plan for using surveys, since surveys provide the capability of collecting opinions of a specific population (Story and Tait, 2019). For the purpose of evaluation, we plan to build the model in the form of a website, where users can easily explore and interact with elements of the model and their descriptions. We aim to provide tasks related to using the model and run online survey questionnaires, including both open- and close- ended questions.

In our survey questionnaire, we plan to collect feedback, which help complement and improve our model and extend its generalizability. We also plan to collect aspects such as project characteristics such as the software development process (e.g. SAFe, XP, Less, Scrum), product domain (e.g., web domain, embedded software domain, Health domain), and software development team size. Collecting such knowledge will help us better analyze the effect of factors explained in our model, enrich the model and increase its generalizability. We plan to contact participants for our survey through our networks and in online professional groups (e.g., ASD groups in LinkedIn).

Our long-term evaluation plan, is to extend the evaluation in companies. Thus, we plan to assess the use of the model in ASD cases and validate it.

## 6. DISCUSSION

### 6.1 The significance of QR documentation in ASD

Our work supports the importance of documenting QRs in ASD. For instance, it reveals that practitioners identify documenting QRs as important in ensuring quality and clarity, as well as in enforcing process conformance. We also found that the project context and QR type affect the level of QR documentation required in ASD. For instance, the level of documentation needed for projects employing safety is greater than those implementing basic utility software or a Web app. Stettina and Heijstek (2011) reported that more than half of the ASD practitioners in their study perceived documentation as important, although they were not asked specifically about QR documentation. All the practitioners who participated in our study acknowledged that it is important to document QRs.

Interviewees from all three cases stated that documenting QRs contributes to ensuring quality. By specifying QR targets and acceptable levels of quality, documenting QRs helps ensure software quality. We also noted that documenting QRs is perceived to improve practitioners' and customers' understanding of QRs. We believe that this is an important finding since building and maintaining shared understanding among customers is difficult in ASD (Kasauli et al., 2021). Moreover, the lack of QR awareness among ASD customers and practitioners is a challenge in managing QRs in ASD (Behutiye et al., 2020a). In this regard, teams can benefit from documenting QRs.

QRs define the desired properties of software and affect the implementation of interdependent features. The practitioners in our study stated that documenting QRs helps in monitoring changes and improves the traceability of QRs. This is important since a lack of QR traceability is challenging in ASD (Baca et al., 2015; Behutiye et al., 2017). We learned that documenting QRs could be a means of ensuring process conformance regarding specification and documentation in ASD. It helps confirm that practitioners documented QRs properly and did not neglect them.

### 6.2 Factors influencing QR documentation in ASD

We observed that most practitioners compromise QR documentation due to time constraints and, thus, QRs may be underspecified, leading to rework in later phases. Sprint duration influences QR documentation and management practices. We learned that ASD teams tailor their software development process to meet the QR documentation needs of projects, depending on the sprint duration. For instance, ASD teams assign separate theme-oriented sprints (e.g., for implementation or documentation) when the sprint duration is short.

In large-scale distributed ASD settings where there are multiple teams, and each team is responsible for its own tasks, time constraints may be perceived as not affecting QR documentation. Assigning clear responsibilities for QR tasks has been reported as a useful practice to address QR challenges (Behutiye et al., 2020a). The clarity of the tasks and responsibilities may have helped minimize the influence of time constraints on documenting QRs in the specified large-scale distributed ASD setting.

Project contexts determine how QR awareness influences documenting QRs. Some projects value QRs and include activities in the development process to ensure that QRs are documented and managed properly.

However, other projects may not have such a process. In the latter, QR awareness may have a more negative effect on QR documentation. We also noted that in ASD, QR awareness among both practitioners and customers affects documentation. The lack of QR awareness among customers is a challenge in ASD (Behutiye et al., 2020a; Tetmeyer et al., 2015). When considering practitioners, QR awareness may be viewed as a necessity for certain roles, such as project managers and product owners, and deemed less important for developers in some cases. Junior developers may not have an adequate level of QR awareness. In such cases, the recommendation was to include formal documentation practices in the process to improve their understanding of QRs. We observed that in the embedded systems and telecommunications domain, where software is linked with hardware implementations, upfront knowledge of QRs and specifications may be difficult to obtain.

Most of the practitioners stated that communication gaps affect QR documentation in ASD. In some cases, the practitioners presumed that establishing QR specifications in the early phases of a project prevents communication gaps regarding QRs. However, as QRs evolve over time in the development process, there is always potential for communication gaps. In this regard, we believe that QR documentation should be continuously updated. We also noted that open discussion of QRs is encouraged to minimize communication gaps regarding QRs among team members.

## 6.3 Consequences of missing or outdated QR documentation

A widely reported consequence of missing and outdated QR documentation is the accumulation of technical debt, including increased development and maintenance time and system quality degradation. Missing and outdated QR documentation make QRs and their tasks unclear. The extra time spent on clarifying QRs may demotivate ASD practitioners. Moreover, it may also create friction among practitioners, as there will not be a common understanding of QRs. Imprecise QR specifications have led to misinterpretations (Ho et al., 2006). When QRs are not documented and managed properly, the ASD team's software implementations may not meet customers' expectations and thus harm business relationships.

## 6.4 Implications of the model

Our model contributes to the research gap in ASD documentation strategies. It helps advance our understanding of QR documentation in ASD and has the potential to assist QR documentation tasks. The model considers three important factors (i.e., time constraints, QR awareness, and communication gaps regarding QRs among team members) that affect QR documentation in ASD. Focusing on these aspects is helpful in understanding how QR documentation can be affected (e.g., how time constraints may lead to underspecifying and neglecting QRs). The model also presents mitigation strategies (e.g., lightweight artifacts, such as the given-when-then template) that practitioners can adopt to address issues that may occur due to these factors. We discuss the mitigation strategies and present examples. Additionally, the model considers various perspectives, such as prevalence in certain roles and dependency on project characteristics, to provide a more comprehensive understanding and support optimal QR documentation in ASD.

The model uses knowledge gained from interviews, a review of the literature regarding QR management in ASD, and our prior work investigating QR documentation practices in ASD, which we conducted with the three cases in this study. Ho et al. (2016) proposed a model for specifying and testing performance requirements that is suitable for ASD. Unlike their proposal, our model focuses on optimal QR documentation in general, and we do not focus solely on certain QRs. We did not find other models focused on supporting QR documentation in ASD. However, in the ASD literature, some models address the traceability of QRs (Arbain et al., 2017) and aim to integrate usability into ASD (Butt et al., 2014).

Although our model is based on knowledge from industry and the literature, it lacks empirical evaluation. We acknowledge this limitation. In our future work, we aim to evaluate the model and extend the work to derive guidelines that support optimal QR documentation.

## 6.5 Implications for software engineering industry and research

The implications of our work extend to both the software engineering (SE) industry and SE research. For the SE industry, we propose a model to support optimal QR documentation in ASD. The model provides a conceptual representation of factors affecting QR documentation, as well as their effects and some mitigation strategies. Using this model, practitioners (e.g., junior developers and product owners) can identify potential issues that may arise due to limited knowledge of QRs, time constraints, or communication gaps among team members. They can take proactive mitigation actions that will either prevent or minimize the likelihood of such issues. Moreover, practitioners can learn about the significance of documenting QRs (e.g., that documenting QRs helps in ensuring quality and the clarity of tasks, as well as in decision-making) and the consequences of poor specifications and outdated QR documentation (e.g., accumulating technical debt, unhappy customers, rework, incorrect implementations, and lack of a common understanding of QRs). They can also adopt practices and strategies suggested to mitigate issues that may arise from the factors discussed in our paper.

We complement SE research with a study on the importance of documenting QRs in ASD. Researchers can learn about the effect of QR awareness, time constraints, and communication gaps among team members on QR documentation, as well as the perceived consequences of underspecified and outdated QRs in ASD. They can also utilize the model to understand how the factors that we examine affect QR documentation. Moreover, they can use our study as a basis to investigate the topic with other industrial cases. We believe that our findings and the model can be enriched by extending the study with other ASD cases that are operating in varying contexts. For instance, those operating in different product domains (e.g., web, health, Financial systems) and those applying different ASD methods (e.g., companies applying scaled agile framework (SAFE), Large scale Scrum(LESS), and XP).This will be beneficial since additional evidence from industrial cases will help consolidate knowledge on QR documentation in ASD.

## 6.6 Threats to validity

We consulted (Feldt and Magazinius, 2010) validity threats in software engineering and adopted Runeson and Höst's (2009) validity threat classification approach for case studies to discuss the threats in our study. We also present the corresponding mitigation strategies that we took to improve the validity of our study.

**Construct validity**: We communicated the research objective of our study and suggested potential participant roles using the key informant technique to the representatives of the cases. This approach enabled us to collect relevant information from ASD practitioners. We minimized threats from the misinterpretation of concepts and interview questions by describing the research objectives and clarifying concepts and interview questions with the participants. For instance, we clarified that we treat QRs and non-functional requirements as equivalents.

**Internal validity**: To minimize threats to internal validity, we conducted the study with practitioners who have a broad range of experience with ASD. We acknowledge that other factors, which we are unaware of and did not consider, might threaten the internal validity of our results and the model. For instance, considering factors other than time constraint, QR awareness and communication gaps may yield other outcome. However, we took the following measures to address this issue. The first three authors discussed and reviewed the findings and the model during its creation. Representatives from the cases also reviewed and provided feedback on our findings and the model too.

**External validity**: explain the generalizability of a study to a context outside the investigated cases. In our study, we synthesized evidence from three different cases of varying project contexts. For instance, the cases varied in terms of product domains, company sizes, and the ASD method they applied. Our study involved 12 participants with different roles and levels of experience in software engineering and ASD. We observe that some of our findings are shared among the three cases, and believe that they may as well partly extend to cases with similar contexts (e.g. other cases applying Scrum, large scale agile software development, and operating in the embedded, telecommunications domain). For instance, practitioners from the three cases identified QR documentation important for ensuring quality and clarity of QRs, and they determined TD accumulation as consequence of missing and outdated QR documentation. These findings are likely to be reflected in other ASD cases with similar contexts.

The cases in our study have established practices for QR documentation and consider QRs important in their product domain. As reflected in our findings, the participants seem to be aware of the significance of QRs. This may as well be due to the participants' rich experience in software engineering and ASD (median of 16.5 and 10 years respectively). It is possible that conducting the study with other cases, and participants of varying experience in ASD and software engineering, may have a different outcome.

In deriving our model, we considered findings from the study, and the QR documentation practices from the three case and QR documentation and management practices reported in the literature. However, the model's generalizability is still limited since it lacks empirical validation and mainly consider QR documentation practices from the three cases. To enhance the model and its generalizability, we aim to evaluate the model with ASD practitioners and experts working in other cases. We also consider extending our study with other cases. We plan to utilize the knowledge that we obtain from the evaluation and additional case studies to enrich the model and develop it further.

**Reliability**: We applied a protocol to guide our interviews and collected data systematically by recording audio of the interviews to increase the reliability of the study. We acknowledge that the first author performing the data analysis is a threat in the data analysis. However, to minimize the threat of subjective bias, the second and third authors reviewed and discussed the results. In our results, we also provide direct quotes. Additionally, we took triangulation measures by having the representatives of the cases in our study review the findings.

## 7. CONCLUSIONS

In this paper, we examined practitioners' perspectives of the importance of documenting QRs and their perceptions of factors that may affect QR documentation in ASD through a multiple case study of three cases. The three cases have product domains in embedded systems and telecommunications, modeling tools, and telecommunications. We used knowledge driven from this investigation and prior work on QR documentation practices with the aforementioned three cases and a review of the literature of management of QRs in ASD (Behutiye et al., 2020a, 2020c) to propose a model, which is the main contribution of this paper. The model has the ultimate goal of supporting optimal QR documentation in ASD and improved understanding of QR documentation.

We found that practitioners identify documenting QRs as important to ensure software quality, clarify QRs, enforce process conformance on QRs, and enhance decision-making, as well as because QRs influence the implementation of other features. We also found that time constraints, QR awareness, and communication gaps regarding QRs affect QR documentation in ASD. For instance, due to time constraints, QRs may become underspecified, and teams may require additional iterations to handle documentation needs. We found that ASD teams might tailor their documentation practices according to the sprint duration to achieve optimal documentation. Project managers' and product owners' knowledge of QRs is more important than developers' knowledge of QRs. We found that practitioners recommend open discussions regarding QRs to minimize communication gaps. Missing and outdated QR documentation may lead to incurring technical debt, a lack of common understanding regarding QRs, informal QR management processes, and incorrect implementations.

Our study contributes to both the software industry and software engineering research. Software practitioners (e.g., project managers, product owners, and developers) can learn about the significance of documenting QRs in ASD and become informed about how documenting QRs can contribute to ensuring software quality and facilitate software development by clarifying QR tasks and supporting decision-making in ASD. Although the model has not been empirically validated yet, practitioners can possibly benefit from it. For instance, the model identifies factors influencing QR documentation (QR awareness, time constraints, and communication gaps regarding QRs among team members) and the consequences of outdated QR documentation (e.g., lack of common understanding regarding QRs). They can examine how each factors may affect QR documentation based on project characteristics. They can also learn about potential mitigation strategies to prevent documentation issues arising from the aforementioned factors. For software engineering research, our study provides empirical evidence regarding QR documentation in ASD. We also contribute to strategies in agile documentation, which was identified as a gap, by introducing our model. Researchers may utilize the findings to gain additional insight into the research area.

As future work, we aim to extend our work by conducting the study with other cases, as it serves a means of collecting additional evidence on the topic and means for validating our model. We plan to evaluate our model with ASD practitioners and experts. We believe that such empirical evaluation will help us obtain knowledge that may help enhance our model and extend its generalizability. We also encourage other researchers to extend our study with other cases. We also plan to extend our work to derive guidelines that support optimal QR documentation.

## ACKNOWLEDGMENTS

This study is partially funded by the Q-Rapids project, European Union's Horizon 2020 research and innovation funded program under grant agreement N° 732253.

**Appendix: Interview script**

**Introduction and purpose of interview**

*Around 5 Minutes Elapsed 0*

The goal of our interview today is to collect expert opinion feedback regarding the documentation of quality requirements (non-functional requirements) and related practices in ASD. The interview will last approximately 30 minutes. We begin with warm-up questions. Following that, we will ask questions about the documentation of QRs. We would like to collect expert opinion feedback on the documentation of QRs in ASD and on how different factors may affect the documentation of QRs in ASD projects. We would like to record the interview to capture all data more accurately. All data will be anonymized by linking data to interviewee ID and kept confidential. You have the right to stop the recording at any time. This will not affect your work in the company in any way.

**1. WARM-UP QUESTIONS:**

*Less than 5 Minutes Elapsed 5*

Q1: **Could you tell us something about your work experience?**
- ✓ *How long have you been working in the company?*
- ✓ *What is your role in the development/project?*
- ✓ *How many years of experience in software engineering?*
- ✓ *How many years of experience in agile software development?*

## 2. QR DOCUMENTATION QUESTOINS

In this section, we are interested in your feedback as an agile software development expert while answering the questionnaire. These questions are based on evidence reported in the scientific literature of agile requirement studies in different agile projects.

Q2.1 **What is your opinion regarding the documentation of QRs in ASD?**
- ✓ *Do you think it is important to document QRs?*
    - o *If yes, how? Please elaborate.*
    - o *If not, why not? Please elaborate.*

Q2.2 **The findings from scientific literature indicate that time constraints and the short iteration cycles in ASD affect documentation of QRs.**
- ✓ *Do you think time constraints, short iterations of ASD impact the documentation of QRs?*
    - o *If yes, how? Please elaborate.*
    - o *If not, why not? Please elaborate.*

Q2.3 **Do you think limited/inadequate QRs awareness affects the documentation of QRs in ASD?**
- ✓ *If yes, how? Please elaborate.*
- ✓ *If not, why not? Please elaborate.*

Q2.4 **How do you see communication gaps among team members regarding QRs in ASD?**
- ✓ *Do you think it affects documentation of QRs?*
    - o *If yes, how? Please elaborate.*
    - o *If not, why not? Please elaborate.*

Q2.5 **In your opinion, what is the consequence of missing and outdated QR documentation in ASD? Please elaborate.**